%% file: paper_arxiv.tex
\documentclass[twocolumn,floatfix,showkeys]{revtex4-1}
\usepackage{bm}
\usepackage{booktabs}
\usepackage{graphicx} 
\usepackage{grffile}
\usepackage{hhline}
\usepackage{pgfplots}
\usepackage{amsmath}
\usepackage{siunitx}
\usepackage{todonotes}
\usepackage{amsmath}
\usepackage{amsthm}
\usepackage{amssymb}
\usepackage{exscale} % Correct font scaling in formulas
\usepackage{xfrac}

\presetkeys{todonotes}{inline}{}

\newcommand{\figwidth}{8.9cm}

\newcommand{\N}{\ensuremath{\mathbb{N}}}
\newcommand{\T}{\ensuremath{\mathbb{T}}}
\newcommand{\Z}{\ensuremath{\mathbb{Z}}}
\newcommand{\R}{\ensuremath{\mathbb{R}}}
\newcommand{\C}{\ensuremath{\mathbb{C}}}
\renewcommand{\N}{\ensuremath{\mathbb{N}}}
\newcommand{\ti}{\ensuremath{\mathrm{i}}}
\newcommand{\eim}[1]{\ensuremath{\,\mathrm{e}^{-2\pi{\ti} #1}}}
\newcommand{\eip}[1]{\ensuremath{\,\mathrm{e}^{2\pi{\ti} #1}}}
\newcommand{\erf}{\mathrm{erf}}
\newcommand{\erfc}{\mathrm{erfc}}

\newcommand{\x}{{\vec x}}
\newcommand{\bfk}{{\vec k}}

\newcommand{\IM}{\mathcal I_{\vec m}}
\newcommand{\vm}{\vec{\mu}}
\renewcommand{\vr}{\vec r}

\begin{document}
\title{Accelerating the calculation of dipolar interactions in
  particle based simulations with open boundary conditions by means of
  the P$^2$NFFT method}

\author{Rudolf Weeber$^1$}
\email{weeber@icp.uni-stuttgart.de}
\author{Franziska Nestler$^2$}
\author{Florian Weik$^1$}
\author{Michael Pippig$^2$}
\author{Daniel Potts$^2$}
\author{Christian Holm$^1$}

\affiliation{$^1$
Institute for Computational Physics,
University of Stuttgart,
Allmandring 3,
Germany
}
\affiliation{$^2$
Chemnitz University of Technology,
Faculty of Mathematics,
09107 Chemnitz,
Germany}
\date{\today}% It is always \today, today,

\begin{abstract} 
  Magnetic gels are soft elastic materials consisting of magnetic
  particles embedded in a polymer network.  Their shape and elasticity
  can be controlled by an external magnetic field, which gives rise to
  both, engineering and biomedical applications.  Computer simulations
  are a commonly used tool to study these materials. A well-known
  bottleneck of these simulations is the demanding calculation of
  dipolar interactions. Under periodic boundary conditions established
  algorithms are available for doing this, however, at the expense of
  restricting the way in which the gels can deform in an external
  magnetic field. Moreover, the magnetic properties depend on the
  sample shape, ruling out periodic boundary conditions entirely for
  some research questions.  In this article we will employ the
  recently developed dipolar variant of the P$^2$NFFT method that is
  able to calculate dipolar interactions under open boundary
  conditions with an $N \log N$ scaling in the number of particles,
  rather than the expensive $N^2$ scaling of a direct summation of
  pair forces. The dipolar P$^2$NFFT method has been implemented within the ScaFaCoS library. The molecular dynamics software ESPResSo has been extended to make use of the library.

  After a short summary of the method, we will discuss
  its value for studying magnetic soft matter systems. A particular
  focus is put on developing a tuning strategy to reach the best
  performance of the method at a predefined accuracy, and lastly
  applying the method to a magnetic gel model.  Here, adapting to the
  gel's change in shape during the course of a simulation is of
  particular interest.
\end{abstract}


\maketitle

\section{Introduction}
In this article we report on the application of a novel scheme called
particle-particle NFFT method (P$^2$NFFT) to calculate the dipolar interactions in soft magnetic
materials for the case of open boundary conditions. This is a relevant
use case, in particular, for magnetic gels.  These materials consist
of magnetic nanoparticles embedded into a gel
matrix~\cite{barsi96a,varga03a}. This allows to control their shape and
elasticity via external magnetic
fields~\cite{szabo98a,gollwitzer08a,mitsumata11a,odenbach16a} giving
rise to applications such as
actuation~\cite{ramanujan06a,monz08a,zimmermann11a} or biomedical
applications, since the magnetic field controlling their behaviour
does not significantly interact with living matter. A review on magnetic gels can be
found in Ref.\,\cite{weeber18a}, a discussion of their study using particle-based simulations in Ref.\,\cite{weeber18b}.

Simulations of magnetic gels have been performed mostly using
molecular
dynamics~\cite{dudek07a,weeber12a,weeber15a,weeber15c,weeber17a-pre,minina17a}
as well as finite-elements and finite-volume
methods~\cite{stolbov11a,cremer15a,attaran16a,metsch16a,kalina16a}.

Two interesting aspects in the simulation of a magnetic gel are the
calculation of the elasticity, -- and sometimes the structure -- of
the hydrogel matrix, and the calculation of the interactions between
the embedded magnetic particles.  The excluded volume interaction of
the magnetic particles is often 
approximated by hard or soft sphere potentials, while the magnetic interaction
is simplified to a point dipole approximation, with the dipoles
sitting at the particles
center. Hence, the task is to calculate the dipolar interaction energy
\begin{equation}
\label{eqn:dd}
-\frac{\mu_0}{4 \pi} \left[ 3 \frac{\langle{\vec{\mu}_i, \vec{r}_{ij} \rangle \langle{\vec{\mu}_j, \vec{r}_{ij}}\rangle}}{r_{ij}^5} - \frac{\langle{\vec{\mu}_i, \vec{\mu}_j\rangle}}{r_{ij}^3} \right],
\end{equation}
where $\mu_0$ is the magnetic vacuum permittivity, and $\vec{\mu}_i,\vec{\mu}_j\in\R^3$ denote the particles' dipole moments.
Furthermore, we denote by $\vec r_i\in\R^3$ and $\vec r_j\in\R^3$ the positions of the dipoles as well as by
\begin{equation*}
r_{ij}:=\|\vec r_{ij}\|:=\|\vec r_i-\vec r_j\|
\end{equation*}
their distance.

Which of the two parts, elasticity or dipolar interactions, is most
computing intensive, depends on the modeling approach. In those cases,
where the hydrogel or its elasticity is simulated in great detail,
either via the inclusion of explicit polymer chains or via a continuum
model, these calculations dominate the computing time.  The
calculation of the dipolar interactions, on the other hand, is most
relevant for models which model the hydrogel's elasticity in a very
coarse-grained fashion, e.g., by entropic springs connecting the
magnetic particles. These models can then contain a large number of
magnetic particles. In case of Ref.\,\cite{weeber17a-pre}, e.g.,
10\,000 dipolar particles are simulated.

It is important to note that dipolar interactions are long-ranged, as
the interaction scales like $r^{-3}$, where $r$ denotes the distance
between two dipoles. A simple cut-off scheme for reducing the number
of dipole pairs to be considered results in significant errors.  Under
periodic boundary conditions the computational complexity can be
reduced to $O(N \log N)$, where $N$ denotes the number of dipoles in
the system, using particle mesh Ewald methods like the dipolar P$^3$M
method~\cite{CeBaHo11}.  However, the behaviour of a magnetic gel is
also determined by its
shape~\cite{raikher03a,zubarev12a,weeber17a-pre}, due to the
shape-dependent demagnetization energy. Hence, when this effect is
important and needs to be taken into account, open boundary conditions
have to be applied.

While it is possible to calculate the dipolar interactions for an open
system by considering all pairs of particles, the
calculations scale like $O(N^2)$, making the simulation for large
$N$ computationally expensive.  In this contribution, the dipolar
P$^2$NFFT method~\cite{Ne16,HoNePi16} is applied to solve this
problem. By using a Fourier-space based calculation for the long range
part and applying a regularization at the boundaries to obey
the open boundaries, the method achieves an $O(N \log N)$ scaling with
respect to the number of dipoles in the system.

Support for the dipolar P$^2$NFFT method with open, periodic, and
mixed boundary conditions has been added to the ScaFaCoS library~\cite{misc-scafacos,arnold13b}.
The implementation is parallelized by means of MPI.
Support for using the ScaFaCoS library for long range interactions has been added to the ESPResSo molecular dynamics software~\cite{limbach06a,arnold13a}. Furthermore, a tuning
scheme has been developed that allows to find those method parameters
resulting in a given accuracy.  The paper is structured as follows. In
Sec.\,\ref{sec:p2nfft}, a brief overview of the method is presented,
in Sec.\,\ref{sec:tuning} the tuning procedure is described. The
model system, on which the method and tuning are evaluated, is
described in Sec.\,\ref{sec:model}, tuning results as well as timings
and the method's scaling behaviour are then presented in
Sec.\,\ref{sec:params}. Based on this, a heuristic scheme for
selecting method parameters is discussed in
Sec.\ref{sec:heur}. Lastly, we apply the method to a magnetic gel
simulation (Sec.\,\ref{sec:sim}) and conclude with a summary.

\section{Dipolar P$^2$NFFT for open boundary conditions}
\label{sec:p2nfft}

Given a system consisting of $N$ dipoles at positions $\vec r_j\in\R^3$, each possessing a dipole moment $\vec{\mu}_j\in\R^3$,
the energy of this system is given by
\begin{equation}\label{eq:Udd}
U_\mathrm{dd}=-\frac{\mu_0}{4\pi}
\sum_{j=1}^N\sum_{i=1,i\neq j}^N
\frac{\left<\vm_i,\nabla_i\right>\left<\vm_j,\nabla_j\right>}{r_{ij}}.
\end{equation}
Here, $\nabla$ denotes the gradient operator and $\langle \cdot,\cdot \rangle$ represents the Euclidean inner product in $\R^3$.
Further, the single summands are a short form notation of the expressions
\begin{equation*}
\left<\vm_i,\nabla_{\vec x}
\left<\vm_j,\nabla_{\vec y}\frac{1}{\|\vec x-\vec y\|}\right>
\right>
\end{equation*}
evaluated at $\vec x=\vr_i$ and $\vec y=\vr_j$.
Calculating the partial derivatives results in a representation of the single summands as introduced in Eqn.\,\ref{eqn:dd}.

Instead of computing only the overall energy $U_\mathrm{dd}$, one is often interested in computing the single potentials
\begin{equation}\label{eq:pot}
\phi_j := -\frac{\mu_0}{4\pi} \sum_{i=1,i\neq j}^N \frac{\left<\vm_i,\nabla_i\right>}{r_{ij}},
\end{equation}
the fields
\begin{equation}\label{eq:field}
\vec E_j:=-\nabla_j\phi_j,
\end{equation}
the torques
\begin{equation*}
\vec T_j:=\vm_j\times\vec E_j,
\end{equation*}
as well as the acting forces
\begin{equation}\label{eq:force}
\vec F_j:=-\left[\nabla_j\nabla_j^\top\phi_j\right] \vm_j.
\end{equation}

A direct evaluation of the underlying sums of length $N$, cf. Eqn.\,\ref{eq:Udd} or \ref{eq:pot} for instance,
results in an arithmetic complexity of $\mathcal O(N^2)$, which is impractical for large particle systems.
The dipolar P$^2$NFFT method, as introduced in the following, enables an efficient approximation of the introduced quantities
with only $\mathcal O(N\log N)$ arithmetic operations.

In order to derive fast approximation methods in the field of molecular dynamics simulations,
a common approach is to apply the so-called Ewald splitting.
This is especially done in the case that periodic boundary conditions are desired.
Although we are interested in open boundary conditions, we still apply the Ewald splitting in order to separate the interactions
into short range and long range parts, see Sec.~\ref{sec:ewald_splitting}.

The so-called long range part is approximated by a trigonometric polynomial and evaluated based on the well known fast Fourier transform (FFT).
Since the particles have arbitrary positions $\vec r_j\in\R^3$, we make use of the fast Fourier transform for nonequispaced data (NFFT),
which is introduced in Sec.~\ref{sec:nfft}.
Finally, the NFFT based fast summation approach, which combines different NFFT modules, is applied in order to approximate the long range potentials and forces.
We give a short introduction to the fast summation approach in Sec.~\ref{sec:fastsum}.

\subsection{Ewald splitting\label{sec:ewald_splitting}}
The function $f(r)=r^{-1}$, where $r$ denotes a distance between two particles, takes a central role in the field of dipolar interactions,
see Eqns.\,\ref{eq:Udd} and \ref{eq:pot}.
We split the function into two parts via
\begin{equation}\label{eq:splitting}
\frac 1r=\frac{\erf(\alpha r)}{r}+\frac{\erfc(\alpha r)}{r},
\end{equation}
where $\alpha>0$ is the Ewald or splitting parameter, $\erf(\cdot)$ denotes the error function given by
\begin{equation*}
\erf(x):=\frac{2}{\sqrt\pi}\int_0^x \mathrm e^{-t^2}\mathrm dt
\end{equation*}
and $\erfc(\cdot):=1-\erf(\cdot)$ is the complementary error function, see~\cite{Ew21}.
Note that the second summand in Eqn.\,\ref{eq:splitting} is still singular at $r=0$, whereas the first part has a finite limit for $r\to0$, which is given by
\begin{equation*}
\lim_{r\to 0} \frac{\erf(\alpha r)}{r}=\frac{2\alpha}{\sqrt{\pi}}.
\end{equation*}
Furthermore, the $\erf$-term is approaching the value zero very slowly for $r\to\infty$,
whereas the second summand tends to zero exponentially fast.

The potentials can now be computed based on Eqn.\,\ref{eq:pot} by applying the introduced Ewald splitting (Eqn.\,\ref{eq:splitting}) to the single summands with $r:=r_{ij}$.
The included partial derivatives are computed later on.
In case of the potentials we obtain
\begin{equation*}
\phi_j=\phi^\mathrm{short}_j+\phi^\mathrm{long}_j,
\end{equation*}
where we define the short range part
\begin{align}
\phi^\mathrm{short}_j
:=\,&-\frac{\mu_0}{4\pi}  \sum_{i=1,i\neq j}^N \left<\vm_i,\nabla_i\right> \frac{\erfc(\alpha r_{ij})}{r_{ij}} \label{eq:pot_short}
\intertext{and the long range part}
\phi^\mathrm{long}_j
:=\,&-\frac{\mu_0}{4\pi}  \sum_{i=1,i\neq j}^N \left<\vm_i,\nabla_i\right> \frac{\erf(\alpha r_{ij})}{r_{ij}} \label{eq:pot_long1} \\
=\,&-\frac{\mu_0}{4\pi}  \sum_{i=1}^N \left<\vm_i,\nabla_i\right> \frac{\erf(\alpha r_{ij})}{r_{ij}}. \label{eq:pot_long2}
\end{align}
Note that the function $r^{-1}\erf(\alpha r)$ is continuous at $r=0$ and, thus, we allow $r_{ij}=0$.
In addition, the gradient of the radial function vanishes at the origin.
Consequently, the contribution for $i=j$ is equal to zero, i.e., no self-potential is computed.
In contrast, in order to compute the fields, as defined in Eqn.\,\ref{eq:field}, we apply a further differential operator
to the expressions given in Eqns.\,\ref{eq:pot_short} and \ref{eq:pot_long1}.
Thus, we compute a self-field given by
\begin{equation*}
\vec E^\mathrm{self}_j=\frac{\mu_0}{4\pi}\cdot\frac{4\alpha^3}{3\sqrt\pi}\vm_j.
\end{equation*}
Analogously to the self-potential, the self-force is equal to zero.

\subsection{FFT for nonequispaced data\label{sec:nfft}}
We give a brief introduction to fast Fourier transforms for nonequispaced data (NFFT) in three dimensions, cf.~\cite{duro93, bey95, st97, DuSc}.
The NFFT is going to be applied in order to approximate the long range
interactions in Eqn.~\ref{eq:pot_long2}.
Therefor, the interaction kernel $r^{-1}\erf(\alpha r)$ will be approximated by a trigonometric polynomial.
Finally, a combination of different NFFT modules~\cite{pnfft14} will be used in order to evaluate the long range interactions,
see Secs.\,\ref{sec:fastsum} and \ref{sec:p2nfft_0dp}.

In the following we define for some $\vec m=(m_1,m_2,m_3)\in 2\N^3$ the index set
\begin{equation*}
\IM:=\left(
[-\tfrac{m_1}{2},\tfrac{m_1}{2}) \times\ldots\times
[-\tfrac{m_3}{2},\tfrac{m_3}{2})
\right)\cap\Z^3.
\end{equation*}
The NFFT realizes an efficient computation of the sums
\begin{equation}\label{eq:nfft}
f(\x_j)=\sum_{\bfk\in\IM} \hat f_\bfk \eip{\left<\bfk,\x_j\right>},\; j=1,\dots,N,
\end{equation}
i.e., the efficient evaluation of a trigonometric polynomial $f$ at given nonequispaced nodes $\x_j\in[-\sfrac 12,\sfrac12)^3$.
Note that the components of the vector $\vec m$ represent the number of Fourier coefficients present in each single dimension.
In the context of the dipolar P$^2$NFFT method, $|\vec m|=m_1m_2m_3$ is the total number of Fourier coefficients 
used for the approximation of the function $r^{-1}\erf(\alpha r)$.
Later in this article, see Section~\ref{sec:params}, we will restrict our considerations to meshes of size $\vec m=(m,m,m)$, i.e.,
we use the same number of mesh points $m\in2\N$ in each single dimension.

The basic idea of the NFFT is explained as follows.
We simply apply the ordinary inverse FFT on the given equispaced mesh in Fourier space.
Afterward, the values of $f$ in the nonequispaced nodes $\x_j$ are recovered based on the computed equispaced data via a so called window function.

In other words, the function values $f(\x_j)$ are finally approximated via
\begin{equation}\label{eq:nfft_conv}
f(\x_j)\approx \sum_{\vec l\in\IM} g_{\vec l}\;\tilde\varphi\left(\x_j-\tfrac{\vec l}{\vec m}\right),
\end{equation}
where we set $\tfrac{\vec l}{\vec m}:=\left(\tfrac{l_1}{m_1},\tfrac{l_2}{m_2},\tfrac{l_3}{m_3}\right)\in\R^3$.
The function $\tilde\varphi$ is in general the 1-periodization of a compactly supported window function $\varphi$ with small support,
for example a B-spline.
Thus, the sums in Eqn.\,\ref{eq:nfft_conv} consist only of a small number of nonzero summands.

We remark that the NFFT method allows the application of other window functions, such as Bessel functions or Gaussians, for instance. Furthermore, a so called oversampling factor may be applied in Eqn.~\ref{eq:nfft_conv}.
For the sake of simplicity we restrict our considerations in the present paper to the above mentioned B-spline window function without oversampling and refer to Ref.\,\cite{Ne14}, for instance, for further discussions.
We denote by $a\in2\N$ the order of the B-spline and refer to $a$ as the assignment order in the following.

The coefficients $g_{\vec l}$ in Eqn.\,\ref{eq:nfft_conv} are obtained by applying a classical inverse FFT on the given regular grid in Fourier space.
In order to correct for the convolution with the window function afterward, the given Fourier coefficients $\hat f_\bfk$ are deconvolved with the Fourier coefficients of the window function before.
Finally, the NFFT basically consists of the following three steps:
\begin{enumerate}
\item Deconvolution in Fourier space:
\begin{equation}
\hat g_\bfk:= \frac{\hat f_\bfk}{c_\bfk(\tilde\varphi)}
\end{equation}
\item Inverse FFT: $\left(\hat g_\bfk\right)_{\bfk\in\IM}
\mapsto \left(g_{\vec l}\right)_{\vec l\in\IM}$.
\item Approximate the function values by evaluating the short sums \ref{eq:nfft_conv}.
\end{enumerate}
The efficient computation of the sums
\begin{equation*}
h(\bfk)=\sum_{j=1}^N f_j \eim{\left<\bfk,\x_j\right>},\;\bfk\in\IM,
\end{equation*}
is realized similarly based on the FFT. The efficient algorithm is widely known as adjoint NFFT.

In addition, the NFFT has been generalized in order to evaluate the gradients $\nabla f(\x_j)\in\C^3$, see~\cite{pippigdiss},
as well as the Hessians $\nabla\nabla^\top f(\x_j)\in\C^{3\times 3}$, as presented in~\cite{HoNePi16}.
We refer to these variants as the gradient NFFT and the Hessian NFFT, respectively.

A further variant is called adjoint gradient NFFT enabling the efficient computation of
\begin{equation}\label{eq:adj_grad_nfft}
h(\bfk):=\sum_{j=1}^N
\left<\vec f_j,\nabla_j\right> \eip{\left<\bfk,\x_j\right>}  \in\C,
\end{equation}
where $\bfk\in\IM$.
For more details we refer to~\cite{HoNePi16}.
The introduced variants of the NFFT are applied in order to compute the considered dipolar interactions efficiently, see Sec.\,\ref{sec:p2nfft_0dp}.

\subsection{Fast summation for radial kernels\label{sec:fastsum}}
We aim to approximate sums of the form
\begin{equation}\label{eq:fastsum_start}
f_j:=\sum_{i=1}^N \left<\vec c_i,\nabla_i\right> K(\|\x_i-\x_j\|) ,\; j=1,\dots,N,
\end{equation}
where $K:[0,L]\to\R$ is a continuously differentiable function,
$\vec c_j\in\C^3$ are some given vectors and the nodes $\x_j$, $j=1,\dots,N$, are supposed to be arbitrarily distributed in a ball of radius $\sfrac L2$,
i.e., we have $\|\x_i-\x_j\|\leq L$ for all $i,j$.

Note that the long range parts of the potentials (Eqn.\,\ref{eq:pot_long2}) are exactly of the form given in Eqn.\,\ref{eq:fastsum_start},
where we have $\vec c_i=\vm_i$ and $K(r)=-\mu_0 (4\pi r)^{-1} \erf(\alpha r)$.

In Sec.\,\ref{sec:nfft} we introduced the NFFT algorithm, enabling an efficient evaluation of trigonometric polynomials (Eqn.\,\ref{eq:nfft}) at arbitrary nodes $\x_j\in\T^3$.
Now we consider expressions as given in Eqn.\,\ref{eq:fastsum_start}, which are in general non-periodic.
A second difficulty results from the fact that a separation of the source nodes $\x_i$ and the target nodes $\x_j$ is not readily possible, and, hence, a naive evaluation would require $\mathcal O(N^2)$ arithmetic operations.
To overcome these two problems, we extend the non-periodic kernel function $K$ to a smooth periodic function, which we realize based on the so called regularization technique.
Next, we approximate the resulting periodic function by a trigonometric polynomial, which finally enables an efficient evaluation of Eqn.\,\ref{eq:fastsum_start} via separating the source and target nodes.
This ansatz is closely related to the classical NFFT based fast summation approach~\cite{post02}.

If we are able to calculate the derivatives of the function $K$, we may construct a regularized function
$K_\mathrm{R}:[-\sfrac h2,\sfrac h2]^3\to\R$ defined via
\begin{equation}
K_\mathrm{R}(\x):=
\begin{cases}
K(\|\x\|) &: \|\x\|\leq L,\\
K_\mathrm{B}(\|\x\|) &: L<\|\vec x\|\leq\sfrac h2 ,\\
K_\mathrm{B}(\sfrac h2) &: \text{else},
\end{cases}
\end{equation}
with some $h\geq2L$.
The function $K_\mathrm{B}:[L,\sfrac h2]\to\R$ is constructed such that
the values of the first $p-1$ derivatives coincide with those of the kernel function $K$ at $x=L$
and that the first $p-1$ derivatives vanish at $x=\sfrac h2$.
The constant continuation with value $K_\mathrm{B}(\sfrac h2)$ makes the function $K_\mathrm{R}$ smooth on the cube $[-\sfrac h2,\sfrac h2]^3$.
The polynomial $K_\mathrm{B}$, which fulfills the given interpolation conditions, is computed via the so called two point Taylor interpolation, see~\cite{FeSt03,NePiPo13}.

Note that there are no further restrictions to the period $h>0$, except for $h\geq2L$.
In order to describe the size of the regularization domain relative to the size of the whole interval, we introduce the variable $\epsilon\in[0,\sfrac12)$ and write
\begin{equation*}
L=h(\sfrac 12-\epsilon) \iff \epsilon=\frac{h-2L}{2h}.
\end{equation*}

By construction, the $h$-periodic continuation (regarding all three dimensions) of the regularization $K_\mathrm{R}$ is smooth.
Thus, it is well approximated by a three-dimensional trigonometric polynomial
\begin{equation}
K_\mathrm{R}(\x)\approx\sum_{\bfk\in\IM} \hat b_\bfk \eip{\left<\bfk,h^{-1}\x\right>},
\end{equation}
where $h^{-1}\x\in[-\sfrac12,\sfrac 12)^3$.
The Fourier coefficients $\hat b_\bfk$ can be computed via sampling the regularized function $K_\mathrm{R}$ at equispaced nodes and applying the FFT.

Since $\|\x_i-\x_j\|\leq L$ we obtain
\begin{equation}\label{eq:outer}
f_j\approx\sum_{\bfk\in\IM}\hat b_\bfk S_\bfk \eim{\left<\bfk,h^{-1}\x_j\right>},
\end{equation}
where we define
\begin{equation}\label{eq:Sk}
S_\bfk:= \sum_{i=1}^N \left<\vec c_i,\nabla_i\right>\eip{\left<\bfk,h^{-1}\x_i\right>} .
\end{equation}
An efficient approximation of the values $f_j$, as defined in Eqn.\,\ref{eq:fastsum_start}, is now possible as follows.
The sums $S_\bfk$, $\bfk\in\IM$, are evaluated via the adjoint gradient NFFT, cf. Eqn.\,\ref{eq:adj_grad_nfft}, followed by a simple multiplication with the coefficients $\hat b_\bfk$,
and, finally, the sums in Eqn.\,\ref{eq:outer} are computed via the NFFT, cf. Eqn.\,\ref{eq:nfft}.

\subsection{The P$^2$NFFT method \label{sec:p2nfft_0dp}}
In the following we assume that the particles are located in a ball of radius $\sfrac L2$ with $L>0$.
Since the complementary error function tends to zero exponentially fast,
we may reduce the computational effort needed in order to compute the short range potentials (Eqn.\,\ref{eq:pot_short})
by only considering distances $r_{ij}\leq r_{\mathrm{c}}$ with some appropriate near field cutoff radius $r_{\mathrm{c}}<L$.
Of course, such a near field radius only exists in the case that the splitting parameter $\alpha$ is also reasonably chosen.
The same applies to the short range parts of the fields and the forces, which are defined analogously containing further partial derivatives.

Provided that $\|\vec r_j\|\leq\sfrac L2$ for all $j=1,\dots,N$,
we can simply apply the presented NFFT based fast summation approach in order to compute the long range parts of the potentials (Eqn.\,\ref{eq:pot_long2}),
i.e., we set $\vec c_i=\vm_i$ and $K(r)=-\mu_0 (4\pi r)^{-1} \erf(\alpha r)$ in Eqn.\,\ref{eq:fastsum_start}.
We start with approximating the sums in Eqn.\,\ref{eq:Sk} based on the adjoint gradient NFFT, cf. Eqn.\,\ref{eq:adj_grad_nfft}.
Afterward, the sums in Eqn.\,\ref{eq:outer} are approximated via the NFFT, cf. Eqn.\,\ref{eq:nfft}.

In an analog manner, the long range parts of the fields and the forces may be approximated.
Whereas the long range potentials are obtained by applying the NFFT to sums in Eqn.\,\ref{eq:outer}, the long range parts of the fields and the forces are obtained via
\begin{equation*}
\vec E^\mathrm{long}_j =-\nabla_j f_j
\quad\text{ and }\quad
\vec F^\mathrm{long}_j =-\left[\nabla_j\nabla_j^\top f_j\right] \vm_j,
\end{equation*}
respectively.
Here, $f_j$ are approximately given via Eqn.\,\ref{eq:outer}, with $\vec c_i=\vm_i$ and $K(r)=-\mu_0 (4\pi r)^{-1} \erf(\alpha r)$.
Instead of the NFFT we hereby apply the gradient NFFT and the Hessian NFFT, respectively.

\subsubsection*{Error control}
The error behaviour of dipolar particle mesh methods is already well studied regarding fully periodic boundary conditions,
cf.~\cite{WaHo01,CeBaLeHo08,CeBaHo11}. This allows for a fast tuning of the method parameters such that a given accuracy can be reached.
In contrast, there are no error estimation formulae for the presented method under mixed periodic and open boundary conditions.
For open boundary conditions, however, the exact results are easily calculable by means of direct summation. The method parameters of P$^2$NFFT can then be tuned by numercially minimizing the error of the P$^2$NFFT calculations, depending on the method parameters. This is explained in Sec.\,\ref{sec:tuning}.

Recent results indicate that for charge-charge systems, the parameters can be tuned heuristically based on the error estimates known for 3d-periodic constraints~\cite{NePiPo13}. This aproach may be transferrable to the dipole-dipole case in the future, particularly for 1d and 2d periodic geometries, where the exact result cannot be easily calculated.

\section{Model system}
\label{sec:model}

As this paper is focussed on the calculation of dipolar interactions, we use a
dipolar hard sphere system for illustrating the tuning procedure and timing of the method. I.e., soft repulsion between the coatings of magnetic particles as well as the elasticity of the hydrogel matrix in a magnetic gel is not included, see Sec.\,\ref{sec:sim} for the simulation of an actual magnetic gel system.

A fluid of dipolar hard spheres is characterized by two dimensionless quantities. First, the volume fraction 
\begin{equation}
\label{eqn:phi}
\phi =\frac16 \frac{N \pi \sigma^3}{V},
\end{equation}
measuring the ratio of the volume covered by the hard spheres of diameter $\sigma$, and the total simulation volume $V$. 
Second, 
\begin{equation}
\label{eqn:lambda}
\lambda=\frac{\mu_0 \mu^2}{4 \pi \sigma^3 kT}
\end{equation}
is the ratio of the maximum dipolar energy per particle of two touching particles (Eqn.\,\ref{eqn:dd}), divided by the thermal energy $kT$.
The number of particles $N$ and their diameter $\sigma$ can be chosen arbitrarily in a system with periodic boundaries.
In the non-periodic case discussed here, however, the number of particles $N$ is an additional parameter, since it controls the relative influence of bulk and surface properties.
We use reduced units and set the particle diameter $\sigma$ and the thermal energy $kT$ equal to 1.
For the magnetic fluid, experimentally achievable parameters of $\phi=0.05$ and $\lambda=4$ are assumed. Furthermore, model systems with $\lambda \in \{2,4,6,8\}$ at $\phi=0.05$ and for $\phi \in \{0.01,0.05,0.09,0.13,0.17\}$ at $\lambda=4$ are considered for some system sizes.

The test system is generated by randomly placing $N$ particles into the simulation box. Then, overlap between the particles is removed as follows.
A purely repulsive Lennard-Jones potential
\begin{equation}
U(r) = 
\begin{cases}
4\left[ \left(\frac{\sigma}{r} \right)^{12} - \left(\frac{\sigma}{r}\right)^{6}\right] & \mathrm{for} \, r<\sigma, \\
0 & \mathrm{otherwise},
\end{cases}
\end{equation}
is applied to the particles. Then, the interaction energy is minimized by the
steepest descent method. At the end of this procedure, the minimum distance between any two particles in the system is larger or equal to $\sigma$.
Note that this is different from the Weeks-Chandler-Andersen potential, which is cut at $2^{\frac16}\sigma$.
During the tuning and timing of the P$^2$NFFT method, the particles do not move, so the Lennard-Jones interaction is not necessary here. 
We consider system sizes of $5\,000$, $10\,000$, $20\,000$, $40\,000$ $80\,000$, $160\,000$, and $320\,000$ particles.
The orientations of all magnetic dipole moments are chosen randomly. 
By selecting a suitable unit for the electric current, the vacuum permittivity in
Eqn.\,\ref{eqn:dd} can be set to $\frac{\mu_0}{4\pi}=1$.
Inserting the definition of $\lambda$ from Eqn.\ref{eqn:lambda} into the
dipole-dipole interaction (Eqn.~\ref{eqn:dd}), setting the thermal energy $kT=1$ and the particle distance to $\sigma$ yields a value of
\begin{equation}
\mu=\sqrt{ \lambda \sigma^3 }
\end{equation}
for the magnetic moment.

\section{Tuning procedure}
\label{sec:tuning}

At this point, there are no error estimation formulae for the method. However, in contrast to periodic ones, for open boundary conditions the exact result can be obtained from direct summation.
The error for a given set of parameters is then obtained by comparing the P$^2$NFFT result to the exact one.
As error measure we use the average between the root mean square force error and root mean square torque error
\begin{equation}
\label{eqn:error}
\Delta = 
\frac12 \left( \sqrt{\sum\limits_{i=1}^N \vert \vec{F}_i -\vec{F}_i^{\mathrm{exact}}}|^2
+\sqrt{\sum\limits_{i=1}^N |\vec{T}_i -\vec{T}_i^{\mathrm{exact}}}|^2
\right),
\end{equation}
where $N$ denotes the number of particles, and $\vec{F}_i$ and $\vec{T}_i$ denote the force and torque on particle $i$. Exact results are obtained from direct summation.

When applying the method in simulations, the set of parameters has to be found which gives the fastest computation time at the desired level of accuracy. The parameter space is high-dimensional, as it consists of number the of mesh points $m$, the Ewald splitting parameter $\alpha$, the real-space cutoff for calculating short range interactions $r_{\mathrm{c}}$, the assignment order $a$ and the parameter $\epsilon$ controlling the regularization at the boundaries. Therefore not all possible combinations of parameters can be tested. 
We developed a tuning procedure, which consists of two phases. 
First, the set of parameters giving the fastest calculation which still fulfill the accuracy requirement are generated for all combinations of mesh sizes $m$ and assignment orders $a$ to be considered.
In the second phase, the calculations for these parameter sets are timed on the desired number of processor cores, and the parameter set providing the fastest calculation is selected.

Let us now examine the first phase in more detail.
It is important to note that while all parameters can affect the accuracy of the result, the Ewald splitting parameter $\alpha$ and the parameter $\epsilon$ controlling the regularization do not influence the run-time of the algorithm.
Hence in the first step, for a given number of mesh points, assignment order, and real space cutoff, we select the Ewald splitting parameter $\alpha$ and the regularization parameter $\epsilon$ which minimize the error (Eqn.\,\ref{eqn:error}) by means of a numerical minimization using the L\_BFGS\_B method from the SciPy package (\url{www.scipy.org}).
The optimization is constrained: for the Ewald splitting parameter $\alpha$, values between $0.4/r_{\mathrm{c}}$ and $5/r_{\mathrm{c}}$ are considered, where $r_{\mathrm{c}}$ is the real space cutoff. The regularization parameter $\epsilon$ can adopt values between $0.001$ and $0.15$.
The optimization is then repeated for different real space cutoffs. 
For a fixed number of mesh points and assignment order, the smallest real space cutoff which allows the accuracy target to be met results in the fastest simulations.
It is determined by bisecting the interval between the minimum and maximum real space cutoff considered (five and 20 particle diameters ($\sigma$), respectively).
Finally, this procedure is performed for all combinations of the number of mesh points and assignment order, which are to be  considered. As the results do not depend on each other, they can be executed in parallel.

Once viable parameter sets are collected, the calculation is timed on the target
architecture with the desired number of processor cores using these parameter
sets obtained in the previous step, and the fastest one is selected for production simulations.
In practice, both, the root-mean-square force and torque error, and the timing results slightly depend on the configuration of particles in the system. When tuning for an accuracy of $10^{-4}$, we therefore averaged over four configurations. Timing is conducted for eight configurations.

\section{Scaling behaviour and method parameters}
\label{sec:params}
For the purpose of obtaining the scaling behaviour of the method and studying the optimal values of its parameters, tuning was carried out on systems of 5\,000 to 320\,000 particles. Assignment orders $a$ of two to five were considered. For the number of mesh points, candidates were evaluated which are multiples of small prime factors, as this results in faster Fourier transforms.
The full list is
\begin{equation}
\label{eqn:mesh-list}
\begin{split}
	m = & \quad \{ 64,80,96,104,112,120,128, \\
            & \quad 136,144,152,160,176,192, \\
	    & \quad 224,256,288,320,352,384, \\
     	    & \quad 416,448,480,512\}.
\end{split}
\end{equation}
Mesh sizes below $60 (N/10000)^{\frac13}$ were not considered. Furthermore, a time limit of 16 hours on four processor cores was applied to the tuning of any individual combination of number of mesh points and assignment order. Parameter sets which could not be tuned in that time will not provide good simulation performance, in the first place. Applying this limitation, even for the largest system containing 320\,000 particles, 39 viable parameter sets were obtained, with the number of mesh points ranging from 192 to 512.
As mentioned in Sec.\,\ref{sec:tuning}, the sets of parameters that satisfy the accuracy requirements have to be timed to pick the fastest one on a given architecture. Timings were run on 1, 2, 4, 8, and 16 cores on a node containing two Intel Xenon E2630 processors with 8 cores, each. 
\begin{figure}
\centering
\includegraphics[angle=270,width=\figwidth]{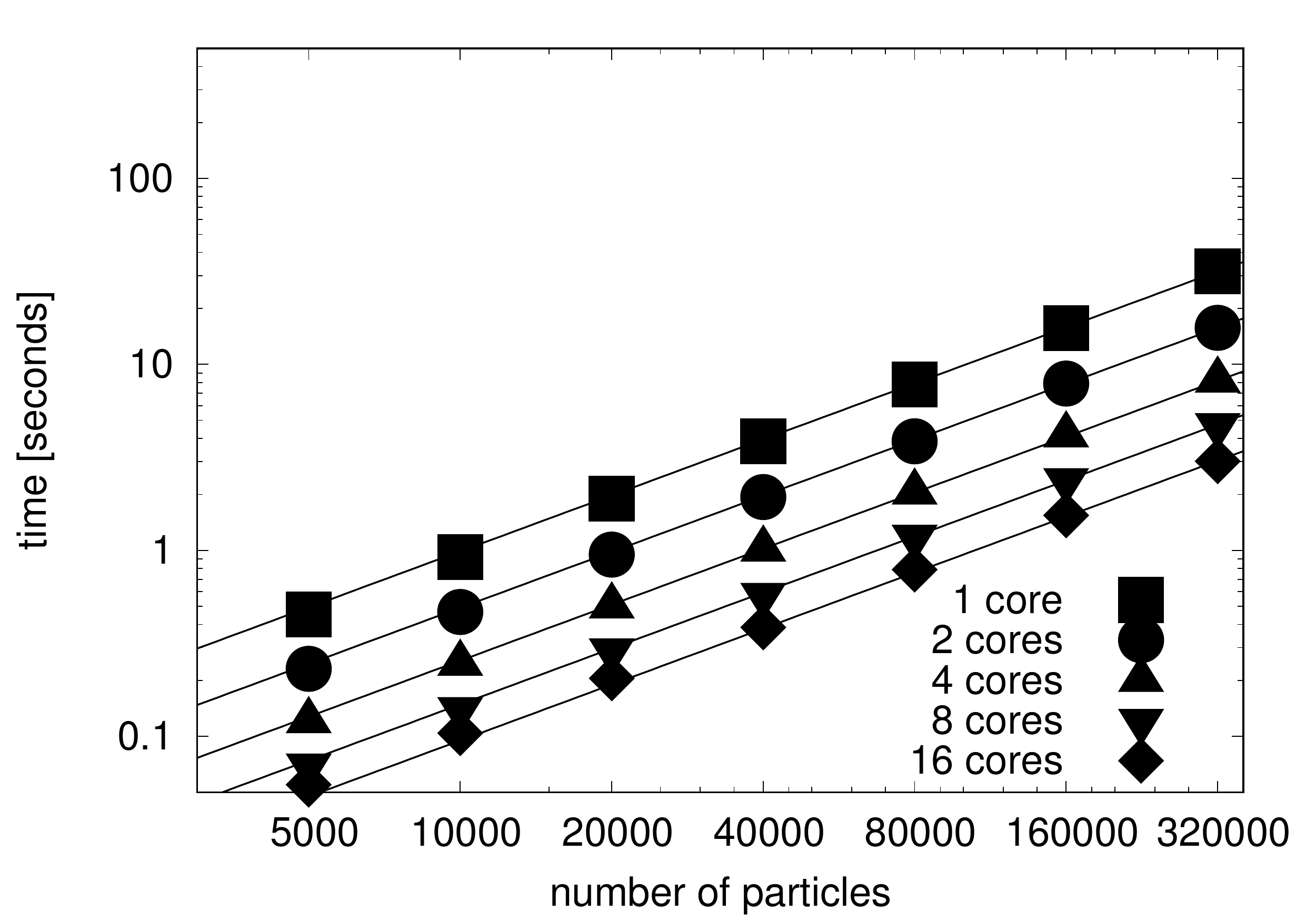}
\caption{\label{fig:strong-scaling}
Strong scaling for calculations on 5\,000 to 320\,000 particles performed on one
to 16 cores. It can be seen that the scaling is close to linear over the complete range of system sizes}
\end{figure}
 In Fig.\,\ref{fig:strong-scaling}, results for the strong scaling are shown. I.e., we plot the time for a calculation versus the number of particles $N$. Data is shown for different numbers of processor cores. It can be seen that the computing time scales almost linearly in the number of particles. The logarithmic component is not significant when increasing the number of particles by a factor of $64$ (5\,000 to 320\,000 particles).
\begin{figure}
\centering
\includegraphics[angle=270,width=\figwidth]{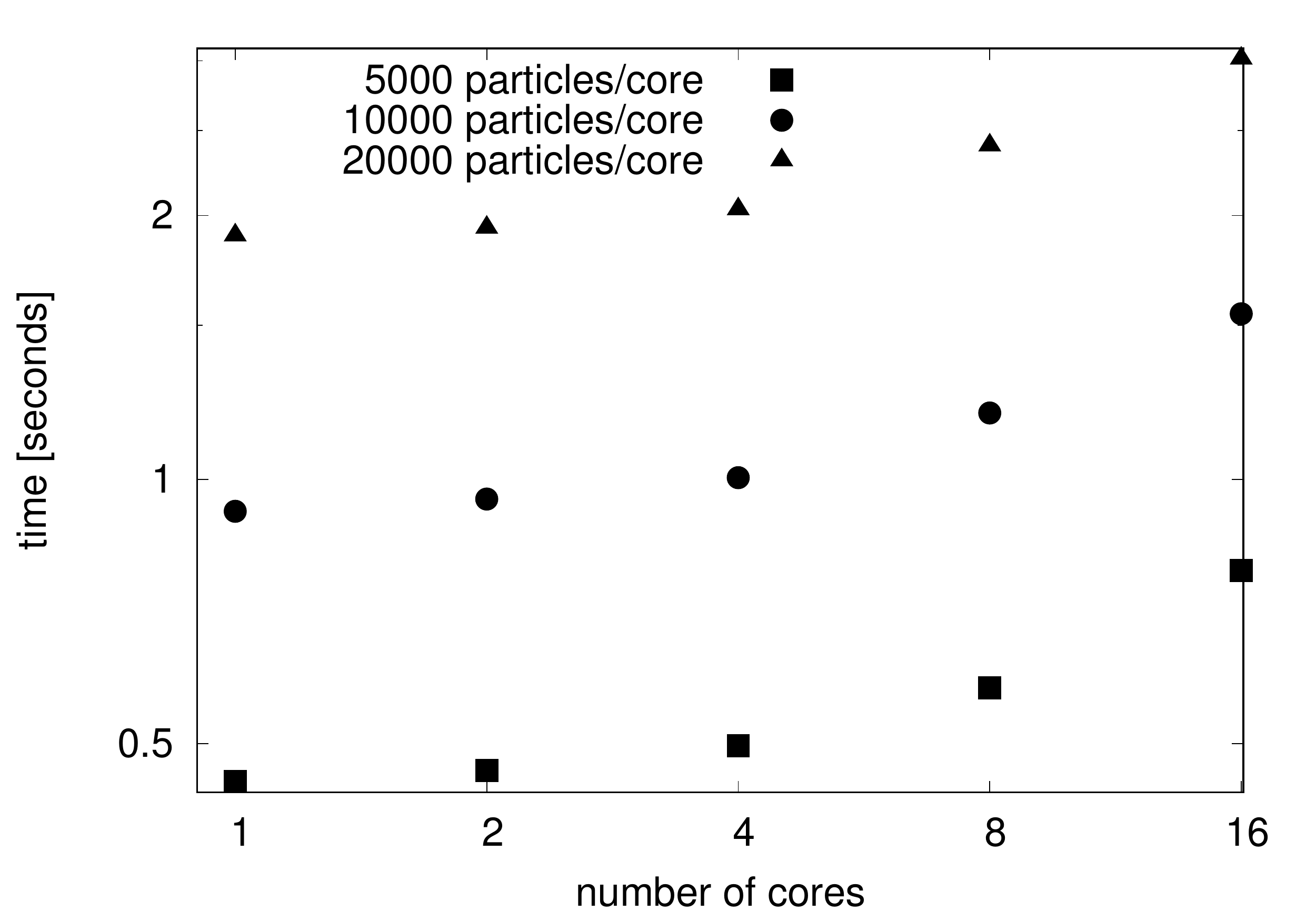}
\caption{\label{fig:weak-scaling}
Weak scaling for calculations on 5\,000, 10\,000 and 20\,000 particles per core for one to 16 cores. The computation time increases with the number of cores, in particular, for 8 and 16 cores. This might be caused by the large amount of data being communicated between the cores.}
\end{figure}

Results for weak scaling are shown in Fig.\,\ref{fig:weak-scaling}. Here, the computing time for a fixed number of particles per processor core is shown versus the number of processor cores. It can be seen that up to four cores, the loss in efficiency is small. For eight and 16 cores, however, the calculations take significantly more time.
This might be related to the rather high communication volume caused by a large real space cutoff and a large FFT grid.

\begin{figure}
\centering
\includegraphics[angle=270,width=\figwidth]{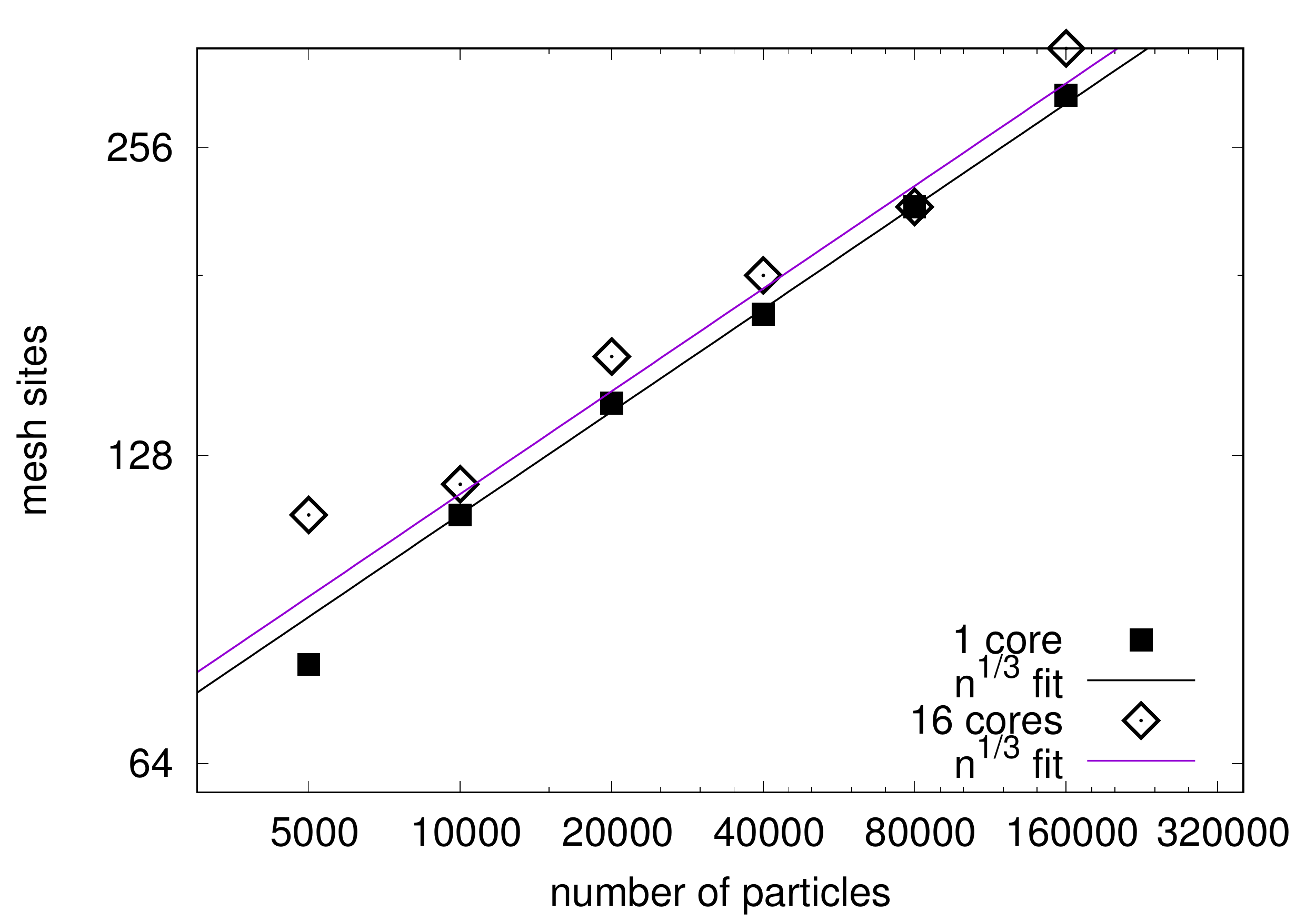}
\caption{\label{fig:mesh-sites}
Number of mesh sites $m$ in each Cartesian direction versus the number of particles $N$. Results are shown for calculations on one and 16 cores, respectively. It can be seen that $m$ scales like $N^{\frac13}$, resulting in an approximately constant ratio of mesh sites per particle $m^3/N$ in the simulation box.
The fluctuations are, in part, caused because for some parameters, an assignment order of $10$ rather than $8$ yielded the fastest calculations.
}
\end{figure}
\begin{table*}
\input{tab-strong-scaling.inc}

\caption{\label{tbl:strong-scaling}
Timings and method parameters for the fastest calculations meeting the accuracy target of $10^{-4}$ on one and 16 cores.
}
\end{table*}

Let us now turn to the parameter sets obtained by the tuning procedure.
The timings and associated method parameters for one and 16 cores and the full range of system sizes can be found in Tbl.\,\ref{tbl:strong-scaling}. It is worth mentioning that in many cases, several sets of method parameters lead to very similar computation times. Often, the five most favorable sets of parameters result in computation time differences of only a few percent. This is due to trade-offs, e.g. between a higher assignment order or number of mesh points on the one hand and a lower real-space cutoff on the other hand.
In Fig.\,\ref{fig:mesh-sites}, the number of mesh sites per direction $m$ is plotted against the number of particles $N$. Results are shown for calculations on one and 16 cores, respectively. They can be fitted with $m\sim N^{\frac13}$. This implies that the ratio of mesh sites per particle $m^3/N$ stays roughly constant for all system sizes. 
Because for some cases an assignment order of $10$ rather than $8$ yielded the
fastest calculations, resulting in a slightly smaller real space cutoff,
we observe some fluctuations of the timing data around the fit in Fig.~\ref{fig:mesh-sites}.
For the dipolar interaction parameter $\lambda=4$ and volume fraction
$\phi=0.05$, we also find that the real space cutoff has to be tuned to $\approx 10\sigma \pm 1.5\sigma$ for all system sizes and number of processor cores. Furthermore, the Ewald splitting parameter scales roughly as $\alpha \sim 1/r_{\mathrm{c}}$, as long as only the system size and the number of processor cores is changed.

\begin{figure}
\centering
\includegraphics[angle=270,width=\figwidth]{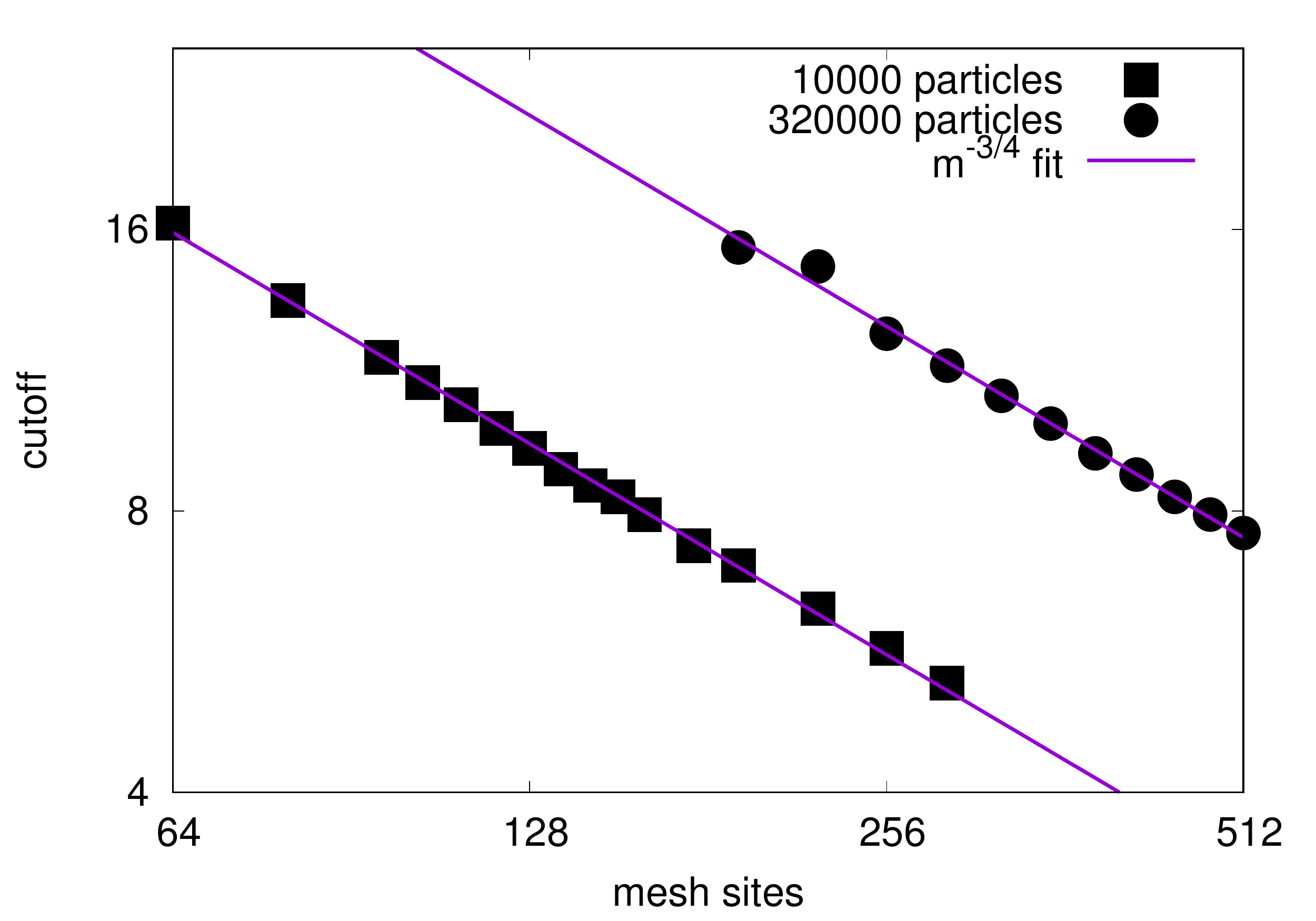}
\caption{\label{fig:cutoff}
Real space cutoff, $r_{\mathrm{c}}$, versus the mesh size per Cartesian direction, $m$, for the parameter sets resulting in the fastest calculation while meeting the accuracy requirement for 10\,000 and 320\,000 particles. The results can be fitted via $r_{\mathrm{c}}\sim m^{-0.75}$.
}
\end{figure}
When choosing the method parameters, a trade-off is made between reducing the effort of the real and Fourier space parts, respectively. The computation time of the real space part scales like $O(r_{\mathrm{c}}^3$), where $r_{\mathrm{c}}$ is the real-space cutoff, whereas the Fourier space part scales like $O(m^3 \log m)$. In Fig.\,\ref{fig:cutoff}, we show the cutoff versus the number of mesh points per Cartesian direction, for the fastest calculations meeting the accuracy target. 
It can be seen that the cutoff scales approximately as $r_{\mathrm{c}}\sim m^{-0.75}$.

\begin{table}
\input{tab-params-vs-rho.inc}
\caption{\label{tbl:params-vs-rho}
Optimal calculation times and associated method parameters for various volume fractions $\phi$ (Eqn.\,\ref{eqn:phi}) for a system of 10\,000 particles with a dipolar interaction parameter $\lambda=4$ simulated on a single cpu core.
}
\end{table}
\begin{table}
\input{tab-params-vs-lambda.inc}
\caption{\label{tbl:params-vs-lambda}
Optimal calculation times and associated method parameters for various dipolar interaction strengths $\lambda$ (Eqn.\,\ref{eqn:lambda}) for a system of 10\,000 particles with a volume fraction $\phi=0.05$ simulated on a single cpu core.
}
\end{table}
Let us finally examine the influence of the strength of the dipolar interactions
$\lambda$ (Eqn.\,\ref{eqn:lambda}) and the volume fraction $\phi$
(Eqn.\,\ref{eqn:phi}) on the optimal calculation time and corresponding method
parameters. Results for a system containing 10\,000 particles with a dipolar
interaction parameter of $\lambda=4$ and for different densities can be found in
Tbl.\,\ref{tbl:params-vs-rho}. It can be seen that the calculation time
increases with increasing density. While the calculations for low densities are
carried out largely in real space, the effort is shifted to the long range
calculations in Fourier space for higher densities. It is worth mentioning that
the effort for the short range calculations also increases with density. This
effort is proportional to the number of particles $n_{\mathrm{c}}$ in a sphere
of radius $r_{\mathrm{c}}$, where $r_{\mathrm{c}}$ is the short range cutoff, we
therefore have
\begin{equation}
n_{\mathrm{c}} \sim \phi r_{\mathrm{c}}^3.
\end{equation}
The ratio between $n_{\mathrm{c}}$ at $\phi=0.17$ and $\phi=0.01$ is approximately 1.83.

Results for varying dipolar interaction parameters $\lambda$ are presented in Tbl.\,\ref{tbl:params-vs-lambda}.  The volume fraction is kept constant at $\phi=0.05$.
While varying $\lambda$ changes the strength of the dipolar interaction, we keep the target accuracy constant at $10^{-4}$. This is justified, because the relative importance of interactions in a soft matter system is measured by their relative strength compared to the thermal energy $kT$. Hence, the required accuracy is determined by this energy scale rather than by the strength of a particular interaction.
From the table, it can be seen that the computation time increases with
increasing interaction strength, both, due to an increase of the short range
cutoff and the number of mesh points.
So, in contrast to the case of a varying volume fraction, the relative importance of short and long range calculations does not change strongly. The increase in computation time is likely mostly due to the error being proportional to $\lambda$.

\section{Tuning heuristics}
\label{sec:heur}

The full tuning procedure described in Sec.\,\ref{sec:tuning} requires a very high computational effort. To make the use of the P$^2$NFFT method practical in simulations, a faster approach is needed.
If tabulated tuning results are available for a similar system, this can be done by extrapolating.
This is possible based on the $m\sim N^{\frac13}$ scaling of the number of mesh points $m$ with the number of particles $N$ (Fig.\,\ref{fig:mesh-sites}). Moreover, while in some cases an assignment order of $a=10$ produced slightly faster calculations, the difference to $a=8$ is not large. 
Based on these observations, an extensive tuning only has to be carried out for a single number of particles. From those tuning results, the number of mesh points for other system sizes can be extrapolated.
We tested this scheme as follows. The number of mesh points ($m_0=112$) for $N_0=10\,000$ particles,  a dipolar interaction parameter of $\lambda=4$, and a volume fraction of $\phi=0.05$ was taken as basis. For all system sizes $N$, considered ($5\,000$ to $320\,000$ particles), the number of mesh points was obtained as
\begin{equation}
m_N =m_0 \left( \frac{N}{N_0} \right)^{\frac{1}{3}}.
\end{equation}
Based on this estimate, the closest $m_N$ was chosen from the list of mesh sizes considered (Eqn.\,\ref{eqn:mesh-list}).
The timing for this $m$ and an assignment order of $a=8$ was then compared against the fastest one obtained for the same system size in the full tuning. 
For $\lambda=4$ and $\phi=0.05$, and for all system sizes above $10\,000$ particles, the computation time using the estimated $m$ was found to be within five percent of the optimal one obtained in the full tuning for the respective system size.
The scheme can even be used for varying volume fractions $\phi$ and dipolar interaction parameters $\lambda$. Then, however, not only the mesh constant from Eqn.\,\ref{eqn:mesh-list} closest to the extrapolated one, but the three closest ones should be considered. In that case, the calculation time was within 15 percent of the optimal one even for the systems with different volume fractions and interaction parameters (Tbl.\,\ref{tbl:params-vs-rho} and Tbl.\,\ref{tbl:params-vs-lambda}) and for the systems with less than 10\,000 particles discussed in the previous section.

In practice, after the number of mesh points has been estimated, only a tuning of the Ewald splitting parameter $\alpha$ and the regularization parameter $\epsilon$ has to be run, which  typically takes less than an hour. This is a low effort compared to a typical soft matter simulation, which runs for millions of time steps.

If no tabulated tuning results are available for extrapolation, it is also possible to employ a simplified version of the full tuning which is based on the assumption that the real space cutoff decreases monotoneously with increasing mesh size. This is applied in the next section.

\section{Application to magnetic gels}
\label{sec:sim}
\begin{figure}
\includegraphics[width=\figwidth]{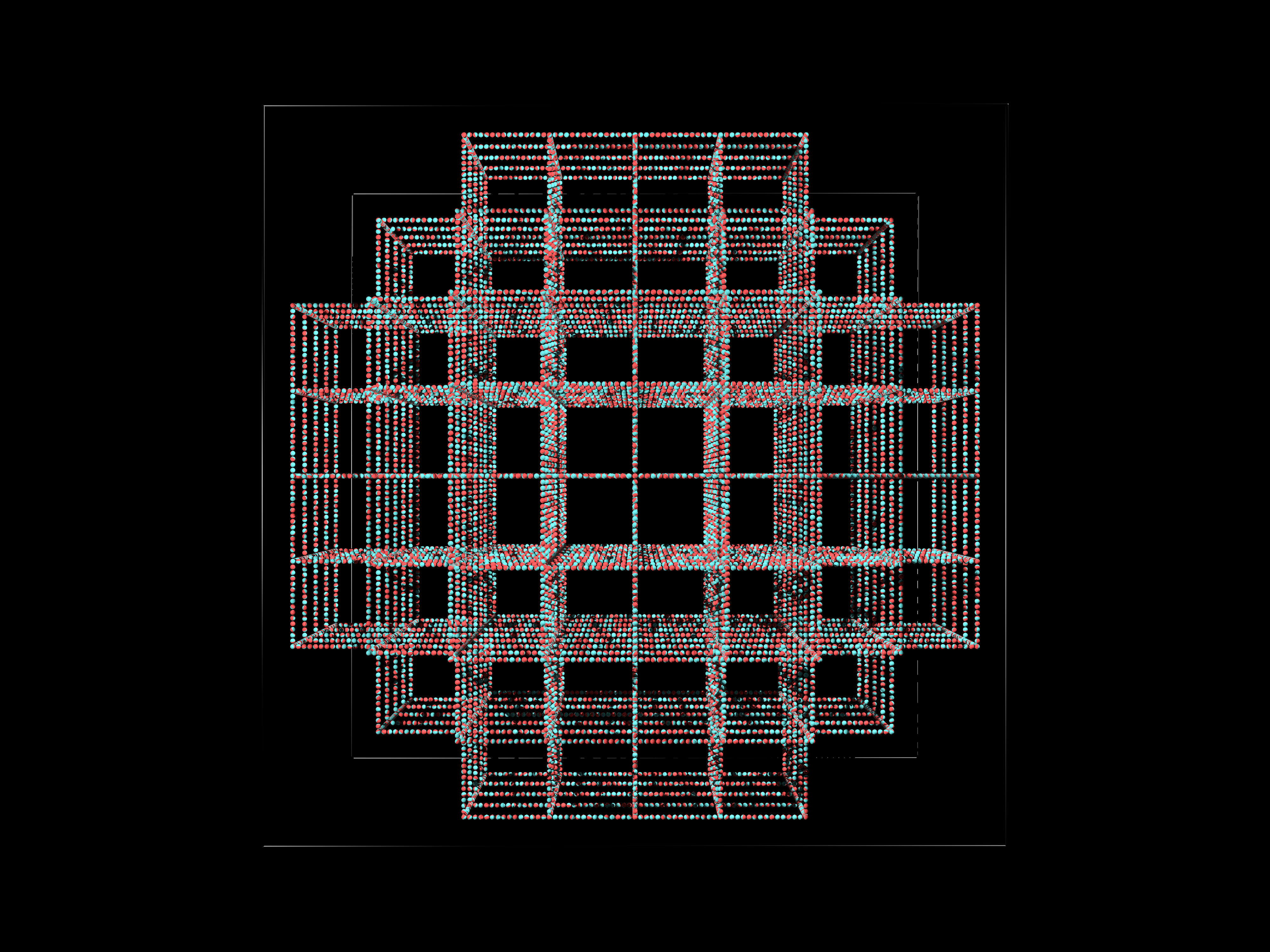}
\caption{\label{fig:demo-init}
Initial configuration of a simple model ferrogel. It consists of a number of flexible chains of magnetic particles. In each node of the network, the ends of six chains are attached. 
}
\end{figure}

\begin{figure*}
\includegraphics[width=\figwidth]{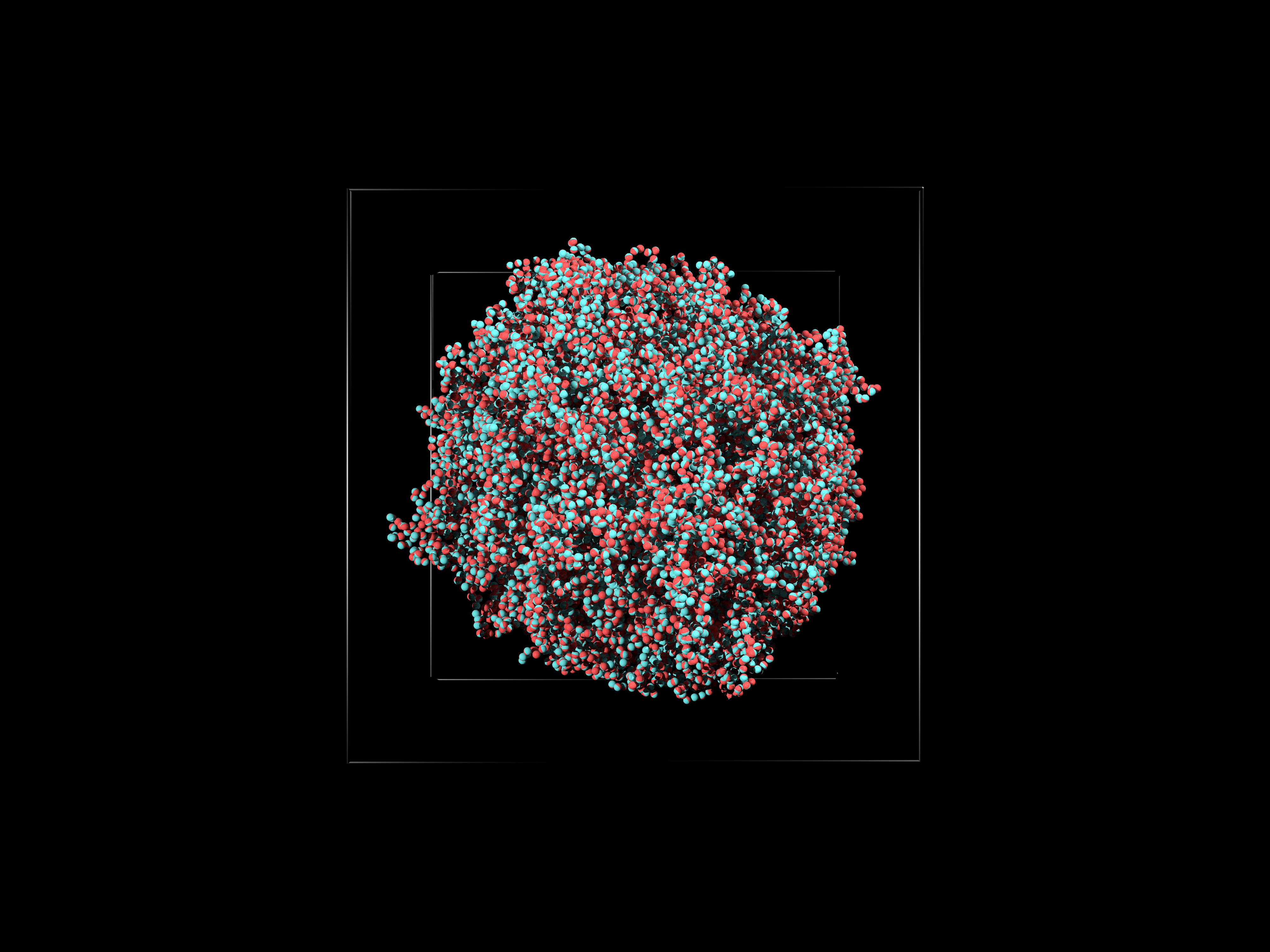}
\includegraphics[width=\figwidth]{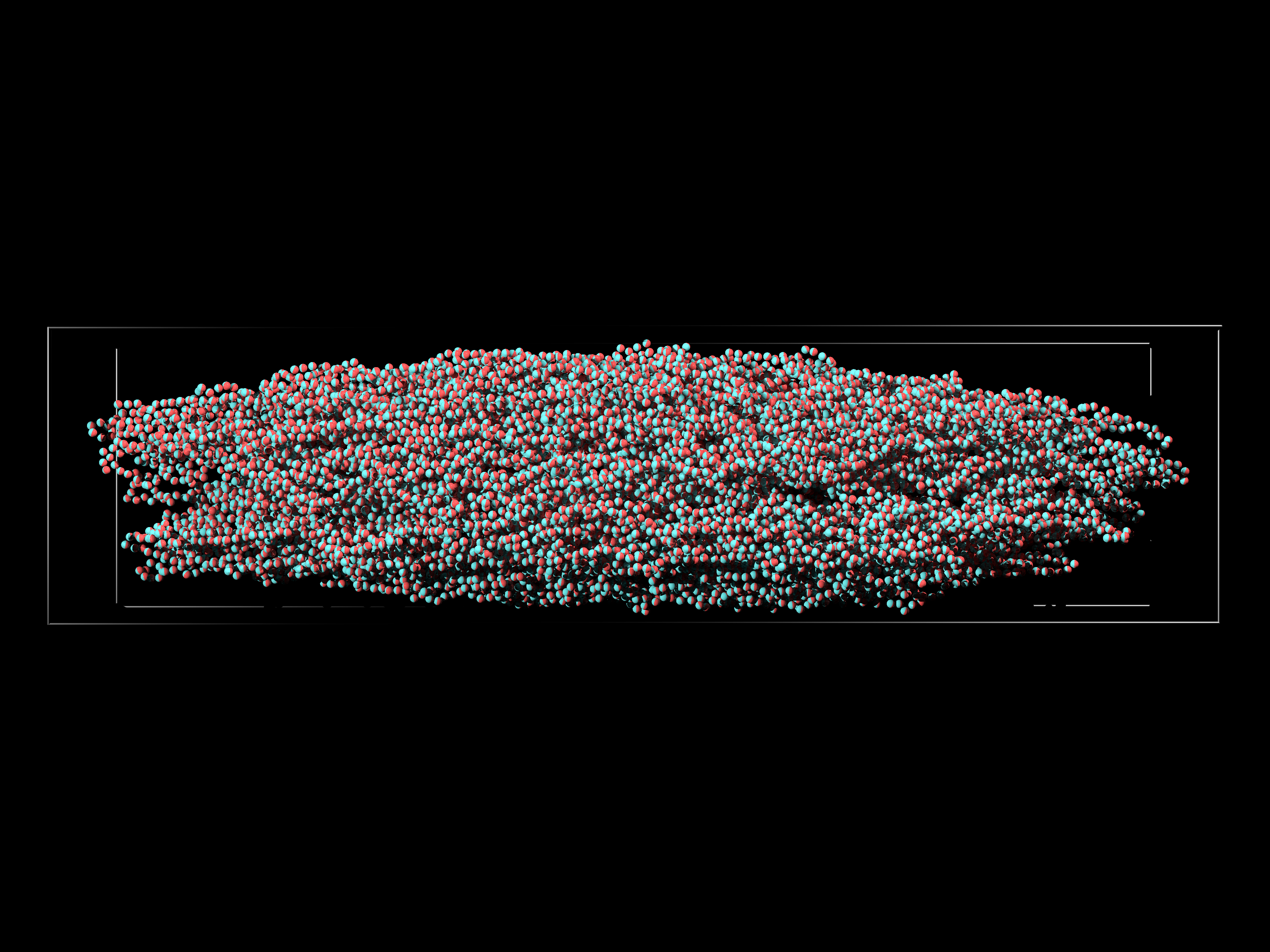}
\caption{\label{fig:demo-result}
Snapshot of the model ferrogel at the end of the simulation. \textbf{Left:} no external field is applied, the gel has a roughly spherical shape. \textbf{Right}: A field of $H=10$ is applied along the $x$-axis. The gel elongates parallel to the field direction and shrinks in the perpendicular one.}
\end{figure*}

Let us now consider the simulation of a magnetic gel. Since these materials can deform quite strongly in an external magnetic field, and since the simulation is run under open boundary conditions, special care has to be taken.
Since the P$^2$NFFT method uses a mesh covering the simulation box, the box
geometry has to be adapted, when the gel changes its shape. A too large
simulation box, where a lot of the mesh does not contain particles, slows down
the calculation.
On the other hand, the sample should also not leave the simulation box. 
Moreover, the P$^2$NFFT method parameters need to be adapted, when the simulation box is changed.
Lastly, the local structure of the material changes considerably in an external magnetic field. This can result in the need for a complete re-tuning of the method after an initial reshaping of the gel.

To illustrate how to cope with these requirements, in this section, we describe a sample simulation protocol. The scripts to run this protocol are provided with this article and at \url{http://github.com/RudolfWeeber/scafacos_espressomd_dipoles}. They can be used as a starting point for new projects.
The `dipoles` branch of ScaFaCoS~\cite{misc-scafacos} with the commit hash 
`066f753f0572c7397508231cb4fc9432d5aeaf04` is used. The commit hash for ESPResSo is 
`7a0ec981f24de839689f61781581ce6194ecf51d`.

The three-dimensional gel model used is loosely based on ``Model I`` in Refs.\,\cite{weeber12a,weeber15a}.
Please note that the model used here is intended as an example application for P$^2$NFFT, not as a model suitable for study of ferrogels.
The model gel consists of a network of flexible chains of $14$ particles, each. 
Additional particles are placed in the nodes of the network, so that the ends of six chains are connected to each node.
As depicted in Fig.\,\ref{fig:demo-init}, the initial arrangement is part of a simple cubic lattice. The gel is cut in a spherical shape and any dangling chains at the surface are removed. In total, there are 17957 particles in the system.
Both, node particles and the particles making up the chains, carry a magnetic moment giving a dipolar interaction parameter of $\lambda=2$ (Eqn.\,\ref{eqn:lambda}). The particles also interact via a purely repulsive Lennard-Jones potential (WCA-potential~\cite{weeks71a})
\begin{equation}
\label{eqn:wca}
U(r) =\left \{
\begin{array}{ll}
\displaystyle{ 4\epsilon \left[ \left( \frac{\sigma}{r}
\right)^{12} -\left( \frac{\sigma}{r} \right)^6 \right] +\frac14}
& r<2^{\frac16} \sigma \\
0 & \mathrm{otherwise},
\end{array} \right\},
\end{equation}
with $\epsilon=\sigma=1$. Neighboring  particles in the chains are connected by a harmonic bond
\begin{equation}
U(r) =\frac12 k r^2,
\end{equation}
with $k=200$. 
The particles are thermalized by a Langevin thermostat, which applies random forces and a velocity dependent friction to the particles~\cite{kubo66a}. Thermal energy and friction are set to $kT=\gamma=1$.
During the simulation, both, the particles forming the chains and the particles in the network's nodes can move. The bonds are maintained throughout the simulation.

\paragraph{Setup of the simulation model}
The initial state shown in Fig.\,\ref{fig:demo-init} is easily constructed but is not yet a realistic representation of a gel in thermal equilibrium.
Due to the entropy introduced at thermal energies $kT>0$, the polymer chains coil and the network as a whole shrinks. This initial relaxation ($1.5\cdot 10^6$ time steps of $dt=0.015$) is done without considering the dipolar interactions between the particles, to save computation time. 

\paragraph{Obtaining suitable P$^2$NFFT parameters}
First it is necessary to calculate the exact dipolar forces and torques for the thermalized system via direct summation. This is then used as reference to calculate the error of the P$^2$NFFT calculation for a given set of method parameters.
Then the tuning is run. For the model gel, we use a simplified version of the tuning procedure described in Sec.\,\ref{sec:tuning}, which avoids the complex workflow and makes use of a single script. Rather than tuning for different mesh sizes in parallel, this is done serially. Assuming a monotone relationship between mesh size and real space cutoff (Fig.\,\ref{fig:cutoff}), the highest and lowest candidate mesh sizes $m$ are tuned first to determine the highest and lowest relevant real space cutoffs $r_{\mathrm{c}}$. Further mesh sizes are attempted in ascending order, such that the maximum real space cutoff to consider can be decreased further. Thus, tuning runs for later in the process will take less time.

\paragraph{Running and re-tuning the model}
Based on the P$^2$NFFT parameters obtained by the tuning, the actual simulation -- including dipolar interactions --, is run. This is done once for the case without an external magnetic field applied, and once for a field of $H=10$. 
Every ten time steps ($dt=0.015$), is is checked whether the gel is too close to the boundary of the simulation box. If there is a layer of less than two ore more than six particle diameters ($\sigma$ in Eqn.\,\ref{eqn:wca}) around the gel, the box is reshaped to a boundary of four diameters. To maintain a good accuracy of the P$^2$NFFT calculation, the mesh size $m$ in each Cartesian coordinate is adjusted to recover the mesh density obtained in the tuning process. Without this adjustment, the accuracy of the P$^2$NFFT calculations deteriorate significantly.

Due to the presence of the external field and the dipole-dipole interactions, the gel changes its shape and inner structure. 
So, even with the adjustment of the mesh size as the box is re-shaped, there is no guarantee that the initial choice of P$^2$NFFT parameters are still a good choice.
Hence, the forces and torques obtained by P$^2$NFFT have to be compared to an exact calculation using direct summation during a long simulation run.

When no external magnetic field is applied, and the gel stays roughly spherical (left part of Fig.\,\ref{fig:demo-result}), we found that the accuracy after 100\,000 time steps is still comparable to the accuracy target of the P$^2$NFFT tuning. 
When, on the other hand, the gel is deformed by an external field, the deviations turn out to be significant, making a retuning of the method after 50\,0000 time steps and a simulation of the remaining 50\,000 time steps with the new parameters necessary.
Images of the model gel both, with and without an external field applied, are shown in Fig.\,\ref{fig:demo-result}.

\paragraph{Scripts provided with this article}
The following scripts are provided with this article and can serve as a starting point for new models
\begin{itemize}
\item \textbf{run-test.sh}\\
Run a short test version of the simulation protocol described above
\item \textbf{run-full.sh}\\
Run the full simulation protocol described above. This is used to obtain the samples shown in Fig.\,\ref{fig:demo-result}.
\item \textbf{model.py}\\Model class
\item \textbf{gen-system.py}\\ Sets up a model gel and thermalizes it
\item \textbf{add-reference-forces-torques.py}\\
Adds reference forces and torques to a stored particle configuration by means of direct summation of the dipolar interactions
\item \textbf{tune.py}\\
Tunes the P$^2$NFFT method for a given particle configuration including reference forces and torques
\item \textbf{run.py}\\
Runs the model, calculating dipolar interactions by means of P$^2$NFFT
\item \textbf{get-accuracy.py}\\
Compares the P$^2$NFFT results for dipolar interaction to an exact calculation by means of direct summation
\item \textbf{p2nfft\_common.py}\\
 Support methods for P$^2$NFFT, e.g., for the re-tuning and mesh size adapting
\end{itemize}

\section{Conclusion and outlook}
In this article, we provided a brief overview of the P$^2$NFFT method
for the calculation of dipolar interactions under open boundary
conditions.  The method is based on the fast
Fourier transform for nonequispaced data (NFFT), and yields a scaling of $N \log N$ in the number
of particles, as compared to the $N^2$ scaling of summing up the
dipolar interactions directly.  In order to make the
NFFT algorithms applicable, the involved long ranged non-periodic
functions are periodized based on a polynomial interpolation and
finally approximated by trigonometric polynomials.

Furthermore, we have demonstrated the usefulness of the algorithm to
magnetic soft matter by showing a sample case, namely the application
to a ferrogel simulation. We have developed a tuning procedure based on a comparison to
exact results obtained by direct summation, 
provide a simulation protocol for a magnetic gel model, and explain how
to cope with a strong change of sample shape and structure during the
simulation.

In summary, we have demonstrated that the P$^2$NFFT method is well suited for the simulation
of large magnetic soft matter systems with open boundaries. However, due to the
effort needed for interpolating onto a regular grid, the method will
outperform direct summation for system sizes on the order of 10\,000
particles and above. 
For the application to magnetic gels, large simulations are important, because the material properties in the bulk and surface regions may be different. Hence, the surface to bulk ratio of simulations should be comparable to that in experiments.
In the future, it may be of interest to off-load
part of the calculations, such as the Fourier transforms, onto an
accelerator such as a graphics card. Since Fourier transforms
perform very well on graphics cards, a considerable speed-up can be
expected. 
Lastly, the P$^2$NFFT method is also applicable to systems with mixed periodicity. Then, a different tuning scheme is required, as the exact result can no longer be calculated directly.

\section*{Acknowledgements}
RW, FW, and CH are grateful for financial support from the DFG through the
SPP 1681, SFB 716 and cluster of excellence Simtech EXC 310, and to the BW-Unicluster for computing resources.
FN, DP and CH gratefully acknowledge support by the German Research
Foundation (DFG), project PO 711/12-1 and HO 1108/25-1.

\bibliographystyle{unsrtinit}
\bibliography{references}

\end{document}

%% file: tab-strong-scaling.inc
\begin{tabular}{|r|rrrrrr|rrrrrr|}
\hline
&
\multicolumn{6}{c|}{1 core} &
\multicolumn{6}{c|}{16 cores}\\
\hline
$N$ &
t [s] & $m$ & $a$ & $r_c$ & $\alpha$ & $\epsilon$ &
t [s] & $m$ & $a$ & $r_c$ & $\alpha$ & $\epsilon$ \\
5000 & 0.453 & 80 & 8 & 11.445 & 0.262 & 0.060 & 0.070 & 96 & 8 & 9.922 & 0.308 & 0.051\\
10000 & 0.920 & 112 & 8 & 10.391 & 0.293 & 0.041 & 0.141 & 112 & 8 & 10.391 & 0.293 & 0.041\\
20000 & 1.897 & 144 & 8 & 10.039 & 0.304 & 0.030 & 0.291 & 144 & 10 & 9.570 & 0.322 & 0.029\\
40000 & 3.860 & 176 & 8 & 10.156 & 0.302 & 0.021 & 0.579 & 192 & 8 & 9.453 & 0.330 & 0.018\\
80000 & 7.805 & 224 & 8 & 9.922 & 0.310 & 0.012 & 1.191 & 224 & 8 & 9.922 & 0.310 & 0.012\\
160000 & 15.674 & 288 & 8 & 9.688 & 0.320 & 0.001 & 2.402 & 256 & 8 & 10.625 & 0.287 & 0.006\\
320000 & 31.726 & 352 & 10 & 9.336 & 0.331 & 0.001 & 4.741 & 352 & 10 & 9.336 & 0.331 & 0.001\\
\hline
\end{tabular}

%% file: tab-params-vs-rho.inc
\begin{tabular}{|r|rrrrrr|}
\hline
$\rho$ & $t$ [s] & $m$ & $a$ & $r_c$ & $\alpha$ & $\epsilon$ \\
\hline
0.01 & 0.649 & 96 & 8 & 15.430 & 0.175 & 0.041\\
0.05 & 0.920 & 112 & 8 & 10.391 & 0.293 & 0.041\\
0.09 & 1.045 & 112 & 10 & 8.867 & 0.356 & 0.048\\
0.13 & 1.104 & 112 & 10 & 8.164 & 0.394 & 0.049\\
0.17 & 1.171 & 120 & 10 & 7.344 & 0.446 & 0.049\\
\hline
\end{tabular}

%% file: tab-params-vs-lambda.inc
\begin{tabular}{|r|rrrrrr|}
\hline
$\lambda$ & $t$ [s] & $m$ & $a$ & $r_c$ & $\alpha$ & $\epsilon$ \\
\hline
2.0 & 0.813 & 112 & 8 & 9.453 & 0.312 & 0.036\\
4.0 & 0.920 & 112 & 8 & 10.391 & 0.293 & 0.041\\
6.0 & 1.002 & 112 & 8 & 10.977 & 0.283 & 0.043\\
8.0 & 1.042 & 112 & 10 & 10.742 & 0.296 & 0.048\\
\hline
\end{tabular}